\documentclass[aps,prb,superscriptaddress,reprint]{revtex4-1}
\usepackage{graphicx}
\usepackage{amsmath}
\usepackage{amssymb}
\usepackage{amsfonts}
\usepackage{filecontents}
\usepackage{color}
\usepackage[normalem]{ulem}
\usepackage{soul}
\usepackage{epstopdf}
\usepackage{hyperref}
\usepackage{fancyhdr}
\definecolor{gold}{rgb}{0.85,.66,0}

\begin{document}

\title{Passive stabilization of hole spin qubit using optical Stark effect}

\author{A. J. Ramsay}\email{To whom correspondence should be addressed: ar687@cam.ac.uk.}
\affiliation{Hitachi Cambridge Laboratory, Hitachi Europe Ltd., Cambridge CB3 0HE, United Kingdom}

\date{\today}

\begin{abstract}
The extrinsic dephasing of a hole spin confined to a self-assembled quantum dot is dominated by charge noise acting on an electric-field dependent g-factor. Here we propose the use of the optical Stark effect to reduce the sensitivity of the effective hole Zeeman energy to fluctuations in the local electric-field. Calculations using measured quantum dot parameters are presented, and demonstrate a factor of 10-100 reduction in the extrinsic dephasing. Compared to active stabilization methods, this technique should benefit from reduced experimental complexity.
\end{abstract}

\maketitle

\section{Introduction}

Spins confined to  self-assembled quantum dots are promising optically active qubits.
Compared to the electron, there are strong reasons to believe that the spin of a heavy-hole \cite{Economou_prb,Muller_prb,Fras_prb,Ardelt_ArXiv,Ramsay_prl,Godden_apl} may make a more robust qubit \cite{Heiss_prb,Gerardot_nat}. The main source of dephasing for the electron spin is the contact hyperfine interaction leading to typical dephasing times of $T_2^*\approx 1.7~{\mathrm{ns}}$, without the use of echo pulses \cite{Press_nphoton}. However, for a heavy-hole with p-type Bloch-function, the contact hyperfine interaction is zero, suppressing the effective hyperfine interaction by a factor of about 10. \cite{Fallahi_prl,Chekhovich_prl,Chekhovich_nphys} In addition, the hyperfine interaction is highly anisotropic, and for an in-plane magnetic field, as used in coherent control experiments \cite{DeGreve_nphys,Godden_prl,Greilich_nphoton} the effect of fluctuations in an out-of-plane effective nuclear magnetic field are further suppressed \cite{Fischer_prb}. However, reported measurements of $T_2^*$ from Larmor precession of the hole-spin (3-20 ns \cite{DeGreve_nphys,Godden_prl,Greilich_nphoton})or coherence population trapping (CPT) ($>$100 ns) \cite{Brunner_sci} vary considerably. The main source of extrinsic dephasing was attributed \cite{DeGreve_nphys,Greilich_nphoton} to charge noise acting on the E-field sensitive in-plane g-factor \cite{Godden_prb,Prechtel_prb,Jovanov_prb}, a view supported by recent studies of spin noise \cite{Kuhlmann_nphys} and CPT \cite{Houel_prl2014}. Methods to reduce the sensitivity of the hole Zeeman energy to charge noise are therefore desirable.

Recently, a number of experiments have demonstrated active stabilization of the emission frequency of a self-assembled quantum dot \cite{Prechtel_prx,Akopian_ArXiv,Hansom_apl}. These are challenging experiments, where the bandwidth is limited by the signal strength. Another approach is to take advantage of the feedback between the Overhauser field and the optical pumping of the nuclear spin bath. This can result in the trion transition locking to a CW laser \cite{Xu_nat,Sun_prl} or the effective Zeeman energy locking to the repetition rate of a pulsed laser \cite{Greilich_sci,Greilich_prb,Varwig_prb,Varwig_prb2014,Fras_prb2012}. In this work, I draw inspiration from work on Silicon quantum dots, where the qubit energy-splitting is engineered to be insensitive to the electric-field. One approach, proposed in refs. \cite{Shi_prl,Koh_prl} and recently demonstrated \cite{Kim_ArXiv} is to use a hybrid qubit composed of three electron spins in a DQD. A second approach explored is to tune the spin-orbit coupling to make the Larmor frequency of a hole spin qubit insensitive to electric-field \cite{Salfi_ArXiv}.  Yet another approach is to use a the AC-Stark shift induced by a CW microwave field to compensate for charge noise induced shifts in the qubit energy-splitting \cite{Laucht_sciAdv,Morello_talk}. Also note the work of Weiss {\it et al} \cite{Weiss_prl} where the insensitivity of a singlet-triplet qubit to electric and magnetic fields at a `sweet-spot' in an InAs double quantum dot leads to long extrinsic coherence times.

In this article, we propose the use of an AC-Stark shift induced by a CW laser to passively stabilize the hole Zeeman energy against charge noise, and thereby increase the extrinsic dephasing time $T_2^*$. A passive stabilization scheme benefits from experimental simplicity. The bandwidth is not limited by a measurement time. In the following, the principle will be explained, and calculations using typical dot parameters are presented where a factor of 10-100 improvement in $T_2^*$ is shown to be possible.

\section{Concept}

Consider a qubit encoded in the spin state of a heavy-hole confined to an InGaAs/GaAs quantum dot. To enable coherent control, a Voigt geometry is used, where a magnetic field $B$ is applied in the sample plane, and the quantum dot is optically excited along the growth-axis. The energy-splitting of the qubit is given by the hole Zeeman energy, $E_{hZ}= g_h(F)\mu_BB$, where $\mu_B = 57.88~\mathrm{\mu eV.T^{-1}}$ is the Bohr magneton. The in-plane hole g-factor $g(F)$ is a result of light-heavy hole mixing. It varies considerably from dot to dot \cite{Schwan_apl1}, is anisotropic \cite{Schwan_apl2}, and depends on the applied electric-field $F$ \cite{Greilich_nphoton,Godden_prb,Jovanov_prb,Prechtel_prb}. As a result, fluctuations in the electric-field at the quantum dot due to charge noise, $\Delta F$ result in an extrinsic dephasing rate $\Gamma_{F}^*=\frac{\partial E_{hZ}}{\partial F}\Delta F$. To simplify the discussion, for now, only fluctuations in the vertical electric-field are considered. The effects of lateral electric-fields will be discussed in the appendix. Furthermore, the relatively weak E-field dependence of the in-plane electron g-factor \cite{Prechtel_prb}, and interactions with the nuclear spins are neglected.

The g-factor has been measured to be linear with vertical electric-field \cite{Godden_prb,Prechtel_prb,Houel_prl2014}, $g_h(F)=g_h(0)+aF$. In the following, quantum dot parameters measured in refs. \cite{Godden_prl,Godden_prb} are used. There a $T^*_2=15~\mathrm{ns}$ was measured at 4.7~T, for a quantum dot with $a=0.0035~\mathrm{V^{-1}\mu m}$, from which we infer that the E-field fluctuations responsible for the extrinsic dephasing is $\Delta F= 42~\mathrm{mV.\mu m^{-1}}$. Values of $g_h(0)=0.051$ and $g_e=0.46$ are used. Our strategy is to use an AC-Stark shift \cite{Xu_sci,Jundt_prl,Muller_prl2008,Muller_prl2009,Brash_prb,Gerardot_njp} to reduce $\frac{\partial E_{hZ}}{\partial F}$ and hence the sensitivity to charge noise.

\begin{figure}
\begin{center}
\vspace{0.2 cm}
\includegraphics[scale=0.3]{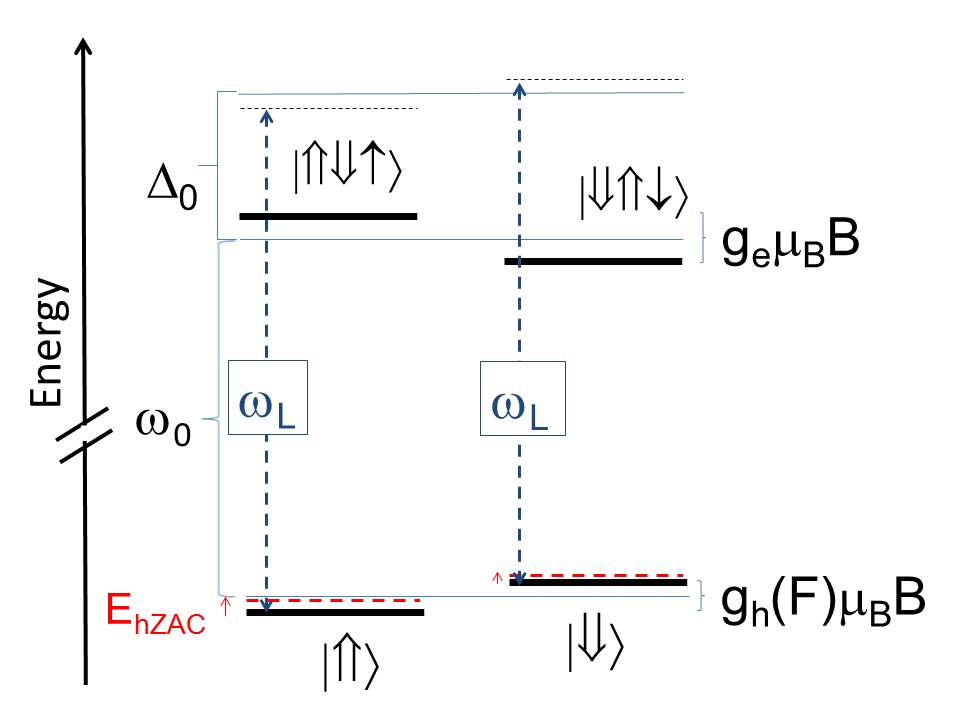}
\vspace{0.2 cm}
\end{center}
\caption{Energy-level diagram of heavy-hole trion system in a Voigt geometry. The hole and trion spin states are aligned parallel and anti-parallel with the magnetic field along the x-direction, and separated by their respective Zeeman energies. The  case of a linearly x-polarized laser positively detuned from both optical transitions is sketched. For a strongly detuned laser, the trion state is not significantly populated and the different AC-Stark shifts on the hole spin states act to reduce the hole Zeeman splitting. For y-polarization, the cross transitions are activated, and the AC-Stark shifts increases the hole Zeeman energy.
}\label{fig:energylevels}
\end{figure}

Figure 1(a) presents an energy level diagram of a 4-level heavy-hole/ positive trion system in a Voigt geometry \cite{Brunner_sci,DeGreve_nphys,Godden_prl,Greilich_nphoton}.  A magnetic field is applied in the sample plane, and the control laser is applied along the growth-axis. The energy eigenstates of the heavy-hole and trion spins are aligned and anti-aligned with the magnetic field. The selection rules are linearly polarized. The case of an linearly x-polarized CW-laser is shown in fig. \ref{fig:energylevels}. A y-polarized laser couples the diagonal transitions: $\vert\Uparrow\rangle \leftrightarrow\vert\Downarrow\Uparrow\downarrow\rangle , ~\vert\Downarrow\rangle\leftrightarrow\vert\Uparrow\Downarrow\uparrow\rangle$.  The laser of Rabi energy $\Omega$ is strongly detuned from all of the optical transitions to avoid populating the trion states. In this weak coupling regime, the laser induces an AC-Stark shift on both of the heavy-hole states, without admixing the states. The difference in the AC-Stark shifts \cite{Ramsay_sst}, results in a shift in the hole Zeeman energy, $\Delta E_{hZAC}^{\pm}$,
\begin{eqnarray}
\Delta E_{hZAC}^{\pm}=\mp\frac{\Omega^2}{4}(\frac{1}{\Delta(F)-E_{\pm Z}}-\frac{1}{\Delta(F)+E_{\pm Z}})
\end{eqnarray}
where $\Omega$ is the Rabi energy. The label $+(-)$ is used for a x(y) linearly polarized laser, respectively. The detuning between the laser, of photon energy $\omega_L$ and the mean of the optical transitions $\omega_0(F)$ is
\begin{eqnarray}
\Delta(F)=\omega_L-\omega_0(F).
\end{eqnarray}
The exciton transition energy $\omega_0(F)$ depends on the electric-field due to an in-built dipole $p=0.4 ~\mathrm{e.nm}$, and an induced electric dipole $\beta=0.02~\mathrm{e.nm.V^{-1}.\mu m}$ as \cite{Fry_prl}
\begin{equation}
\omega_0(F)=\omega_0-pF-\beta F^2.
\end{equation}
$E_{\pm Z}=(g_e\pm g_h(F))\mu_B B$ is the Zeeman splitting of the optical transitions addressed by the laser.  Equation (1) treats the system as two independent two-level atoms, where the Rabi energy is small compared to the laser detuning from both optical transitions: $\Omega\ll \vert \Delta(F)\pm E_{\pm Z}\vert$ .

\begin{figure}
\begin{center}
\vspace{0.2 cm}
\includegraphics[scale=0.9]{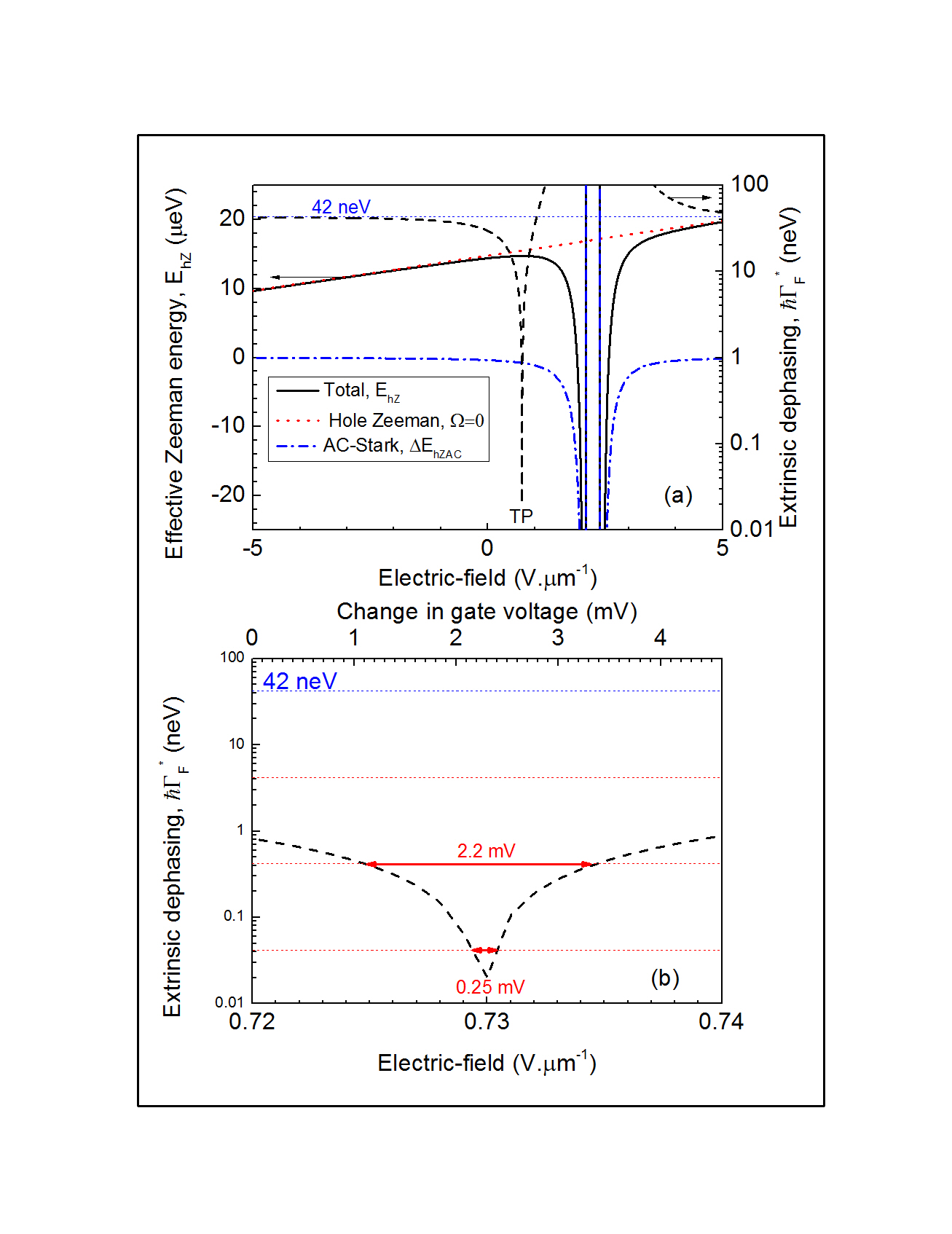}
\vspace{0.2 cm}
\end{center}
\caption{(a) (left-axis) E-field dependence of hole Zeeman energy. x-polarization $(B=5~\mathrm{T}, \omega_L-\omega_0= -1~\mathrm{meV}, \Omega=100~\mathrm{\mu eV})$
(right-axis) At the turning point TP, the hole Zeeman energy is relatively insensitive to E-field, leading to a drop in the extrinsic dephasing rate due to E-field fluctuations by 3 orders of magnitude.
(b) Close-up of turning point. Red-lines indicate factor of (10, 100, 1000) improvement in extrinsic dephasing time. The top-axis indicates change in gate voltage required to change the electric-field, assuming a gate separation of 230~nm \cite{Godden_prb}.}
\label{fig:energyvsF} 
\end{figure}

Figure \ref{fig:energyvsF} illustrates the passive stabilization scheme. The red-trace shows the linear increase of the hole-Zeeman energy with electric-field without an applied laser, at a magnetic field of 5 T. The blue trace presents the AC-Stark shift to the Zeeman energy, for a moderate Rabi energy of $100~\mathrm{\mu eV}$ and a detuning $\Delta_0=\omega_L-\omega_0=-1~\mathrm{meV}$. Due to the quantum confined Stark effect, the optical transitions move through the laser, tuning the AC-Stark effect. The black trace presents the total hole-Zeeman energy $E_{hZ}=E_{hZ}(F)+\Delta E_{hZAC}^{+}$. At an electric-field set-point of $F_0$, there exists a turning-point in the energy where the hole-Zeeman energy is insensitive to fluctuations in the electric-field. The dashed-line provides an estimate of the extrinsic dephasing rate due to E-field fluctuations:
 \begin{equation}
 \Gamma_{F}^*=\vert \frac{\partial E_{hZ}}{\partial F}\Delta F+\frac{1}{2}\frac{\partial^2 E_{hZ}}{\partial F^2}\Delta F^2\vert.
  \end{equation}
  At the `sweet-spot' the dephasing rate drops by about 3 orders of magnitude.

  To assess the robustness of the sweet-spot, fig. 2(b) shows a close-up of the turning point. For a typical device with a  gate separation of 230~nm, the gate voltage needs to be set within 2.2 (0.25) mV of the TP to achieve a factor of 100 (1000) improvement in the extrinsic dephasing time, respectively. Alternatively, the laser photon energy would need to be set within 5.3 (0.61)~$\mathrm{\mu eV}$) of the TP.

    To summarize, the AC-Stark effect can be used to create a turning point in the hole-Zeeman energy versus electric-field where the hole-spin is insensitive to fluctuations in the E-field induced by charge noise.

  \section{Estimate of suppression of dephasing }

\begin{figure}
\begin{center}
\vspace{0.2 cm}
\includegraphics[scale=1.0]{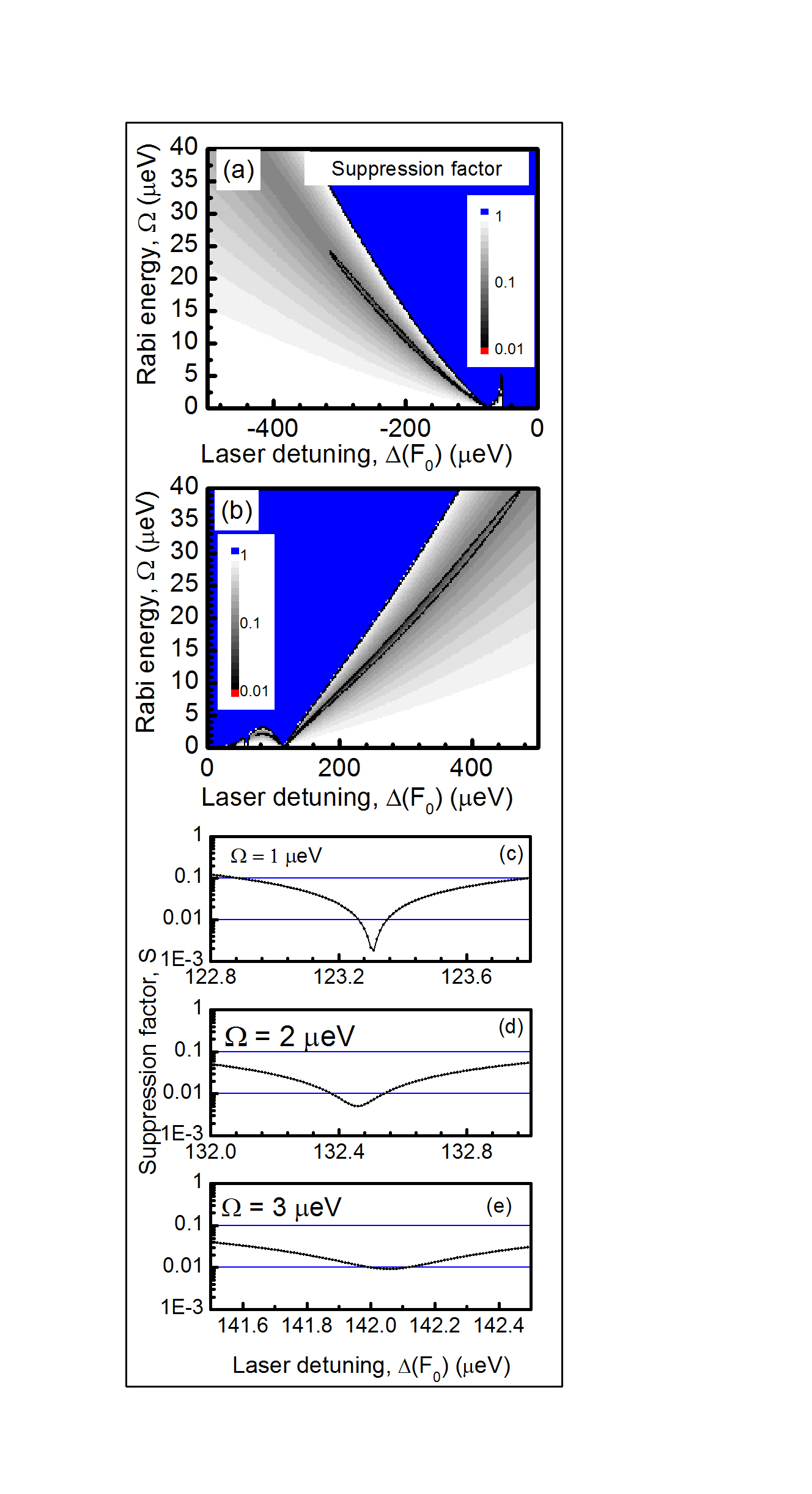}
\vspace{0.2 cm}
\end{center}
\caption{ Grayscale plot of suppression factor of extrinsic dephasing $S=\frac{\Gamma^*}{\Gamma^*(\Omega=0)}$ versus laser detuning and Rabi energy. (a) x-polarization (b) y-polarization (c-e) Cross-sections at fixed Rabi energy $\Omega$ for y-polarization.
B = 5~T. Power fluctuations of $\Delta\Omega^2=\alpha\Omega^2$, $\alpha=0.01$ are used.
}\label{fig:dephasing} 
\end{figure}

However, there is a downside. Applying a laser introduces additional dephasing processes that need to be considered. Firstly, fluctuations in the laser power will introduce additional extrinsic dephasing:
\begin{equation}
 \Gamma_{P}^*=\vert\frac{\partial \Delta E_{hZAC}}{\partial \Omega^2}\Delta\Omega^2\vert.
 \end{equation}
 Secondly, the laser will generate some population in the trion states. When the trion radiatively recombines, to either hole spin state, the hole spin is randomized and this will act as a decoherence process. We estimate an upper limit on this dephasing to be \cite{footnote}:
 \begin{equation}
 \Gamma_{r}^*=\Gamma_R \frac{\Omega^2}{8}\vert\frac{1}{(\Delta(F)+E_{\pm Z})^2}+\frac{1}{(\Delta(F)-E_{\pm Z})^2} \vert,
 \end{equation}
 where $\Gamma_R=1~\mathrm{\mu eV}$ is a typical radiative recombination rate.

To calculate the net dephasing rate, I assume the fluctuations are independent, and add the contributions in quadrature:
\begin{equation}
\Gamma^{*2}=\Gamma^{*2}_F+\Gamma_P^{*2}+\Gamma_r^{*2}.
\end{equation}

 Figure \ref{fig:dephasing}(a,b) present grayscale maps of the suppression factor,
 \begin{equation}
  S=\frac{\Gamma^*(\Omega,\Delta(F_0))}{\Gamma^*(0,\Delta(F_0))}
  \end{equation}
  the ratio of the extrinsic dephasing rate with and without the control laser, against the laser detuning, $\Delta(F_0)$, and Rabi energy $\Omega$. A dark value of $S<1$ indicates an improvement.
  The magnetic field is set to $B=5~\mathrm{T}$, and the electric-field set-point to $ F_0=4~\mathrm{V.\mu m^{-1}}$. The power fluctuations are assumed to be 1\% of total power, which is typical for a laser.
  The black contour indicates the stable regime where an improvement of at least a factor of 10, $S<0.1$, is achieved. The stabilization works better for y-polarization, fig. 3(b), than x-polarization, fig. 3(a). This is a bit counter-intuitive, since the Zeeman splitting between the optical transitions is larger for x-polarization, $E_{+Z}>E_{-Z}$, and the cancelation of the AC-Stark shifts is smaller. However, in the case of y-polarization, the AC-Stark shift adds to the effective hole Zeeman energy, rather than subtracts, resulting in a turning point further from the optical resonance. This reduces the negative impact of $\Gamma_r^*$ due to trion generation. The blue region indicates where the extrinsic dephasing is worse as a result of the control laser, mostly due to the trion generation. The power fluctuations $\Gamma_P^*$ have little influence.

  Figures \ref{fig:dephasing}(c,d,e) present close-ups of the detuning dependence at fixed Rabi energy, near the point where the dephasing is optimized. The inclusion of trion generation reduces the suppression factor that can be achieved, and favors the use of low Rabi energies of a few $\mathrm{\mu eV}$. Increasing the Rabi energy increases the robustness of the stabilization, at the expense of the optimum value of the dephasing rate. At $\Omega=2~\mathrm{\mu eV}$, suppression factors of better than 2-orders of magnitude can be achieved over a detuning range of about 40 neV, equivalent to a change in gate voltage of $16~\mathrm{\mu V}$ for a device of 230-nm gate separation.

Other possible contributions to dephasing introduced by the control laser that are neglected are heating and photo-generation of charge noise. These are discussed in appendix B.

\section{Magnetic field dependence}

\begin{figure}
\begin{center}
\vspace{0.2 cm}
\includegraphics[scale=1.0]{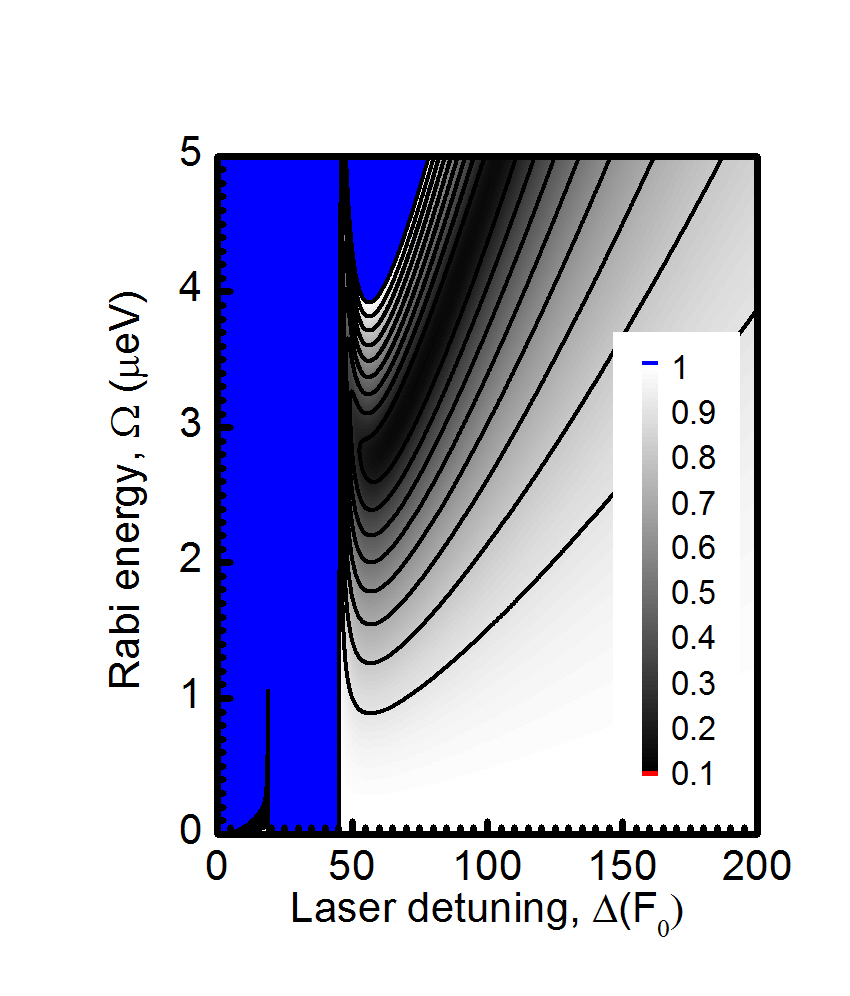}
\vspace{0.2 cm}
\end{center}
\caption{ Grayscale plot of suppression factor against detuning and Rabi energy for magnetic field of $B=1.0~\mathrm{T}$.}
\end{figure}

So far, I have considered a strong magnetic field of 5~T, where the extrinsic dephasing due to fluctuations in the g-factor due to charge noise is more of an issue. Figure 4 presents a calculation of the suppression factor $S$ for a moderate magnetic field of 1~T, for y-polarization. The sweet-spot lies closer to the optical resonance as the strength of the AC-Stark shift to the hole Zeeman splitting, $\Delta E_{hZAC}$ increases with the Zeeman splitting between the optical transitions $E_{-Z}$. Consequently, the gradient in $\Delta E_{hZAC}$ is sharper and the optimum suppression factor is reduced to $S<0.2$, corresponding to an extrinsic dephasing rate of $\Gamma_*<1.7~\mathrm{neV}$, or an extrinsic dephasing time of $T_2^*> 375~\mathrm{ns}$.

The stabilization scheme will work much better in a Faraday geometry, where the magnetic field is parallel to the laser. There a circular polarized laser will couple to only one of the hole spin states, leading to a larger AC-Stark shift for the hole Zeeman energy, and much better performance at low magnetic fields.

\section{Conclusions}

To conclude, the use of a laser induced AC-Stark effect to partially cancel the electric-field dependence of the hole-Zeeman energy and suppress extrinsic dephasing of a quantum dot hole spin due to charge noise is proposed. Calculations that consider additional dephasing induced by the laser, find that for optimized laser detuning and power for a typical quantum dot the extrinsic dephasing can be suppressed by a factor of $>10$. The potential use of the AC-Stark effect to stabilize an optical transition has also been considered. In the case of the hole spin, the optical transition energies are more sensitive to electric-field than the hole-Zeeman energy, and a relatively low power can provide sufficient AC-Stark shift to compensate the hole Zeeman energy. However, in the case of stabilizing an optical transition a much higher Rabi energy, comparable to the laser detuning  is needed, resulting in strong optical pumping, and a high $\Gamma_R^*$.

\section{Acknowledgements}

This work was funded by Hitachi Europe Ltd.

\appendix

\section{Fluctuations in lateral electric-field}

To simplify the discussion, only fluctuations in the vertical electric-field have so far been considered. This approximation holds either if the system is isotropic, or if the system is highly anisotropic. Three factors need to be considered. (i) The electric-field dependence of the in-plane hole g-factor, $g_h(F)$. (ii) The electric-field dependence of the optical transition energy $\omega_0(F)$. (iii) The relative size of the fluctuations in the electric-field.

A typical InGaAs/GaAs self-assembled quantum dot has a truncated square pyramid shape of height 5~nm, and a base of 18~nm \cite{Bruls_apl}, with an Indium rich top \cite{Fry_prl}. Compared to the vertical direction, the quantum dot is relatively symmetric in the sample plane. Therefore the electric-field dependence of the in-plane hole g-factor has the form:
\begin{equation}
g_h(F)=g_h(0)+aF_z+b_{zz}F_z^2+b_{xx}F_x^2+b_{yy}F_y^2
\end{equation}
The g-factor tuning arises from a displacement of the hole wavefunction changing the overlap with the Indium rich regions of the quantum dot \cite{Jovanov_prb,Prechtel_prb}, therefore $b_{zz}> b_{xx},b_{yy}$. In addition, if a vertical electric-field $F_0$ is applied to the quantum dot, $b_{zz}F_0 \gg b_{xx}\delta F_x, b_{yy}\delta F_y$. Hence the effect of a lateral electric-field on the hole g-factor is relatively small. For many quantum dots, the g-factor is linear with applied vertical electric-field \cite{Godden_prb,Prechtel_prb,Jovanov_prb}, and to the best of my knowledge, there are no reports on the lateral-field dependence of the hole g-factor.

If the hole g-factor is insensitive to the lateral electric fields, fluctuations in the optical transition energy $\omega_0$ will dephase the hole-spin due to fluctuations in the AC-Stark shift of the hole Zeeman energy.
Including lateral electric-fields modifies eq. (3):
\begin{equation}
\omega_0(\mathbf{F})=\omega_0-pF_z-\beta_{zz}F_z^2-\beta_{xx}F^2_x-\beta_{yy}F^2_y.
\end{equation}
The contribution to the extrinsic dephasing rate due to lateral x-component of electric-fields can be estimated using eq. (1) as:
\begin{equation}
\frac{\Gamma^*_{Fx}}{\Gamma^*_{Fz}(\Omega=0)}=\frac{\frac{\partial\omega_0}{\partial F_x}}{\frac{\partial\omega_0}{\partial F_z}}\frac{\Delta F_x}{\Delta F_z}\equiv S_x
\end{equation}
where $\Gamma^*_{Fz}(\Omega=0)=42~\mathrm{neV}$ is the extrinsic dephasing rate due to the fluctuations in the hole Zeeman splitting due to fluctuations in the vertical electric-field. At the sweet-spot, this is equal to the extrinsic dephasing rate without the control laser. In the case of isotropic fluctuations in the electric-field, $\Delta F_x=\Delta F_z$, $S_x\approx \frac{\beta_{xx}\Delta F_x}{p+\beta_{zz}F_{z0}}$. Lateral field measurements of the polarizability of InGaAs/GaAs quantum dots give typical values for $\beta_{xx}= 0.03 \mathrm{e.nm.V^{-1}.\mu m}\sim \beta_{zz}$ \cite{Vogel_apl}. Therefore an estimate of the contribution of the lateral fields to the suppression factor at the sweet-spot is $\sqrt{2}S_x=6.5\times 10^{-4}$, using $\Delta F_x=\Delta F=42 \mathrm{mV.\mu m^{-1}}$. This is negligible compared to the suppression factors of $S\sim 10^{-2}$ achieved at the sweet-spot.

A final consideration is the relative sizes of the fluctuations in the lateral and vertical electric-fields. In ref. \cite{Houel_prl}, shifts in the emission energy of quantum dots in a photodiode structure were attributed to charging of defect states at the AlGaAs/GaAs interface. This is typically $\sim 100~\mathrm{nm}$ above the quantum dot. This suggests that the strongest E-field fluctuations in atypical photodiode structure are mostly along the vertical direction.

\section{Heating and photo-generated charge noise}

The optimum Rabi energies are a few $\mathrm{\mu eV}$. This is small compared to Rabi energies of $10-100 ~\mathrm{\mu eV}$ reported in refs. \cite{Xu_sci,Jundt_prl,Muller_prl2008,Muller_prl2009,Brash_prb,Gerardot_njp}, where no evidence of heating was reported. Hence, heating by the stabilization laser is neglected.

In refs. \cite{Jundt_prl,Brash_prb}, under high excitation powers, an additional power dependent blue-shift in the exciton transitions, attributed to the photo-generation of charges within the device structure, was observed. This will result in a power dependent shift in the set-point of the laser detuning needed to stabilize the hole Zeeman energy. More seriously, photo-generated charge noise may be significant, and the charge noise is power dependent, i.e. $\Delta F(\Omega^2)$. The value of $\Delta F$ used in these calculations is inferred from an experiment where the power incident on the sample from the coherent control pulse sequence is about $2~\mathrm{\mu W}$ \cite{Godden_prl}. This is much larger than the 20-nW power needed to achieve a continuous wave Rabi energy of $1~\mathrm{\mu eV}$ \cite{Gerardot_njp} needed to stabilize the hole spin. Therefore, the additional photo-generated charge noise due to the stabilization laser is likely to be small compared to the charge noise introduced by the laser pulses used to control the hole spin.


\bibliographystyle{apsrev}

\begin{thebibliography}{1}
\bibitem{footnote}
For a two-level system, the occupation of the upper-state resulting from a CW-laser detuned by $\Delta$ is $n_T\approx(\Omega/2\Delta)^2$. Then we assume an occupation of $1/2$ in both hole-states.
\end{thebibliography}

\end{document}